\documentclass{emulateapj}

\begin{document}
\title{Velocity Dispersion Profiles of Seven Dwarf Spheroidal Galaxies\footnote{This paper includes data gathered with the 6.5-m Magellan Telescopes at Las Campanas Observatory, Chile, and with the 6.5-m MMT telescope.}}
\shorttitle{dSph Velocity Dispersion Profiles}
\author{Matthew G. Walker\altaffilmark{1}, Mario Mateo\altaffilmark{1}, Edward W. Olszewski\altaffilmark{2}, Oleg Y. Gnedin\altaffilmark{1},\newline Xiao Wang\altaffilmark{3}, Bodhisattva Sen\altaffilmark{4}, and Michael Woodroofe\altaffilmark{4}}
\email{mgwalker@umich.edu}
\altaffiltext{1}{Department of Astronomy, University of Michigan, Ann Arbor}
\altaffiltext{2}{Steward Observatory, The University of Arizona, Tucson, AZ}
\altaffiltext{3}{Department of Mathematics and Statistics, University of Maryland, Baltimore County, Baltimore, MD}
\altaffiltext{4}{Department of Statistics, University of Michigan, Ann Arbor}

\begin{abstract} 
We present stellar velocity dispersion profiles for seven Milky Way dwarf spheroidal (dSph) satellite galaxies.  We have measured 8394 line-of-sight velocities ($\pm 2.5$ km s$^{-1}$) for 6804 stars from high-resolution spectra obtained at the Magellan and MMT telescopes.  We combine these new data with previously published velocities to obtain the largest available kinematic samples, which include more than 5500 dSph members.  All the measured dSphs have stellar velocity dispersion of order 10 km s$^{-1}$ that remains approximately constant with distance from the dSph center, out to and in some cases beyond the radius at which the mean surface brightness falls to the background level.  Assuming dSphs reside within dark matter halos characterized by the NFW density profile, we obtain reasonable fits to the empirical velocity dispersion profiles.  These fits imply that, among the seven dSphs, $M_{vir} \sim 10^{8-9} M_{\odot}$.  The mass enclosed at a radius of 600 pc, the region common to all data sets, ranges from $(2-7)\times 10^7 M_{\odot}$ .  

\end{abstract}

\section{Introduction}

Spanning the absolute magnitude range $-13 \leq M_V \leq -4$ \citep{mateo98,belokurov07short}, dwarf spheroidal (dSph) galaxies represent the extreme end of the galaxy luminosity function.  The Milky Way's (MW's) $\sim 15$ known dSph satellites have managed thus far to avoid and/or to survive interactions that consumed many of their siblings during hierarchical formation of the MW halo.  Recent velocity measurements for several hundred stars per dSph \citep{kleyna02,kleyna03,kleyna04,munoz05short,munoz06short,walker06a,walker06b,simon07} demonstrate that dSph velocity dispersions remain approximately flat with radius.  Analyses that recover $M(r)$ from the Jeans equation conclude that dSphs possess extended and dominant DM halos (e.g., \citealt{lokas02}), unless stellar velocity distributions are highly anisotropic or ongoing tidal disruption invalidates the assumption of equilibrium (e.g., \citealt{kroupa97}).  %\citet{strigari07} find that the enclosed dSphs mass at $r\sim 600$ pc is relatively insensitive to the assumed form of the density profile.  

In this letter we present velocity dispersion profiles for seven dwarf satellites of the Milky Way---Carina, Draco, Fornax, Leo I, Leo II, Sculptor, and Sextans.  Profiles are calculated from new velocity data we have obtained using the Magellan and MMT telescopes \citep{walker07,mateo07}.  After combining with previously published data, samples range in size from $\sim 200$ (Leo II) to $\sim 2000$ (Fornax) member stars per dSph, for a total of 5544 probable members in the seven dSphs.  The new samples more than double the amount of existing dSph kinematic data and allow us to measure velocity dispersion profiles with unprecedented precision.  All the measured dSphs exhibit approximately flat velocity dispersion profiles that can be fit reasonably well under the assumption that dSphs are equilibrium systems embedded within dark matter halos.  We show that if the dark matter halos are characterized by the NFW density profile \citep{navarro96} that results from N-body simulations of cold dark matter, the kinematic data suggest dSphs have $M_{vir}\sim 10^{8-9} M_{\odot}$ and masses of $(2-7) \times 10^7 M_{\odot}$ inside a radius of 600 pc.

\section{Observations \& Data}

Over seven observing runs between 2004 March and 2007 January, we used the Michigan/MIKE Fiber System (MMFS) at the Magellan/Clay 6.5m telescope to obtain 7383 high-resolution echelle spectra of 5793 individual red giant candidates in the dSph galaxies Carina, Fornax, Sculptor and Sextans.  \citet{walker07} describe the uniform data reduction procedure in detail.  MMFS spectra sample the Mg-triplet (MgT) region, spanning $5140-5180$ \AA\ with effective resolution $\sim 0.1$ \AA/pix ($R \sim 20000$).  We measure both stellar velocity and the pseudo-equivalent width of the magnesium-triplet absorption feature.  The latter quantity correlates with surface gravity and helps determine dSph membership.  Comparisons of repeat measurements for more than 1000 stars indicate median measurement errors of $\sim 2.0$ km s$^{-1}$ and $\sim 0.06$ \AA, respectively.  

Additionally we obtained 1183 spectra from 1011 red giant candidates in the dSphs Leo I, Leo II, and Draco using the multi-fiber Hectochelle spectrograph at the MMT 6.5m telescope during three observing runs in March/April of 2005, 2006 and 2007 (see \citealt{mateo07}).  Hectochelle spectra sample the MgT region over 5150-5300 \AA\ with effective resolution 0.01 \AA/pix ($R \sim 25000$).  Repeat measurements for more than 100 stars with Hectochelle indicate velocity errors of $\sim 2.6$ km s$^{-1}$.  

To these new, homogeneous data sets we add 2239 velocities previously published for red giant candidates in the observed dSphs (\textbf{Carina}: \citealt{mateo93,munoz06short}; \textbf{Draco}: \citealt{armandroff95,kleyna02}; \textbf{Fornax}: \citealt{mateo91,walker06a}; \textbf{Leo }I: \citealt{mateo98b,koch07a}; \textbf{Leo II}: \citealt{vogt95,koch07b}; \textbf{Sculptor}: \citealt{westfall06}; \textbf{Sextans}: \citealt{hargreaves94b,kleyna04}).  We correct for zero-point offsets and combine measurements of common stars using the weighted (by measurement errors) mean velocity.  We do not correct for the slopes noted by \citet{walker07} when comparing MgT velocities to those measured from the calcium triplet; because the relevant combined samples are dominated by MMFS data, doing so would have no significant impact on the velocity dispersion profiles.

%The line of sight to each of the targeted dSphs passes through the Milky Way's disk and halo, each of which contains stars with magnitudes and colors satisfying our CMD-based target selection.  We therefore expect each sample to carry a degree of contamination that varies with the density contrast between dSph and foreground, thus increasing with distance from the dSph center.  Fortunately, dSph and foreground populations follow distinct distributions in both velocity and magnesium absorption, allowing us to use both measured quantities to diagnose membership and its dependence on position.  For each dSph observed with MMFS, left panels in Figure \ref{fig:members} display scatterplots of the measured velocities, $V$, and magnesium indices, $W$.  Figure \ref{fig:members} clearly distinguishes loci of foreground stars from those of the Carina, Sculptor and Sextans populations.  The members of these three dSphs cluster into narrow velocity distributions and have weak $W$ relative to foreground stars, which we expect to be K-dwarfs whose strong surface gravity increases their atmospheric opacity \citep{ohman34,thackeray39,cayrel91}.  The failure of the Fornax data to show an obviously distinct distribution of foreground stars is likely due to Fornax's relatively high surface brightness and location near the Galactic anti-center.
 
\begin{figure*}
%  \epsscale{1.2}
  \plotone{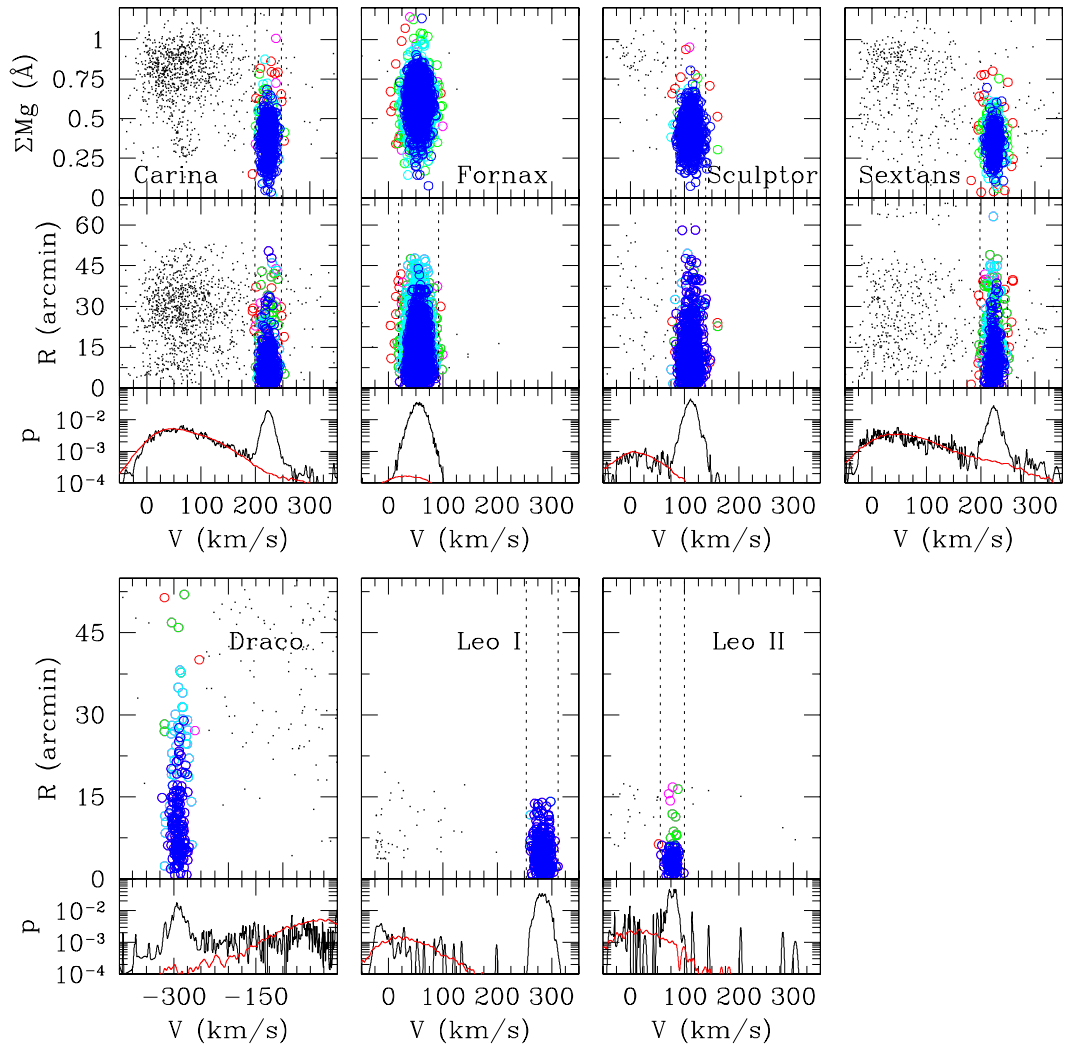}
  \caption{ Mg pseudo-equivalent width, projected distance, and membership fraction vs. velocity for new dSph samples obtained with Magellan/MMFS and MMT/Hectochelle (Mg strength is unavailable for Hectochelle observations).  Marker color indicates the probability that the star is a dSph member.  Black (red; magenta; green; cyan; blue) markers signify $\hat{P}_{dsph} < 0.01$ ($\hat{P}_{dsph} > 0.01; > 0.50; > 0.68; > 0.95; > 0.99$).  Pairs of dotted lines enclose stars that would pass conventional membership tests based on a $3\sigma$ velocity threshold.  Red lines in the bottom sub-panels indicate the interloper distribution predicted by the Besan{\c c}on Milky Way model.  }
  \label{fig:members1}
\end{figure*}

We evaluate the probability, $P_{dsph_i}$, that the $i^{th}$ star is a dSph member using up to three pieces of information---velocity $V_i$, magnesium strength $W_i$ (available only for MMFS samples), and position---\textit{a priori}, stars projected farther from the dSph center are less likely to be members.  We assume 1) the joint distribution of $V$ and $W$ for the members of a given dSph is a bivariate Gaussian; 2) interlopers have magnesium strengths following a univariate Gaussian distribution; and 3) interlopers have a non-Gaussian velocity distribution that we estimate numerically from the Besan{\c c}on Milky Way model \citep{robin03}.  %The probability of obtaining spectroscopic data $Z_i\equiv (V_i,\sigma_{V_i},W_i,\sigma_{W_i})$ for a member star is then  
%\begin{equation}
%  p_{mem}(V_i,\sigma_{V_i},W_i,\sigma_{W_i}) = \frac{\exp \biggl [-\frac{1}{2} \biggl (\frac{[V_i-\langle V \rangle_{mem}]^2}{\sigma_{V_0,mem}^2+\sigma_{V_i}^2}+\frac{[W_i-\langle W \rangle_{mem}]^2}{\sigma_{W_0,mem}^2+\sigma_{W_i}^2} \biggr ) \biggr ]}{2\pi \sqrt{(\sigma_{V_0,mem}^2+\sigma_{V_i}^2)(\sigma_{W_0,mem}^2+\sigma_{W_i}^2)}},
%  \label{eq:pmem}
%\end{equation}
%while the probability of observing $Z_i$ for an interloper is
%\begin{equation}
%  p_{non}(V_i,\sigma_{V_i},W_i,\sigma_{W_i}) = \frac{{p_{bes}}(V_i) \exp \biggl [-\frac{1}{2} \frac{[W_i-\langle W \rangle_{non}]^2}{\sigma_{W_0,non}^2+\sigma_{W_i}^2} \biggr ]}{\sqrt{2\pi(\sigma_{W_0,non}^2+\sigma_{W_i}^2)}}.
%  \label{eq:pnon}
%\end{equation}
%If $p(a)$ is a decreasing function that describes the \textit{a priori} probability that a star at elliptical radius $a_i$ is a dSph member, then given data $Z_i$, the $i^{th}$ star is a dSph member with probability
%\begin{equation}
%  P_{dsph_i}=\frac{p_{mem_i}p(a_i)}{p_{mem_i}p(a_i)+p_{non_i}(1-p(a_i))},
%  \label{eq:emprob}
%\end{equation}
%and the data set $\{Z_i\}_{i=1}^N$ has likelihood quantified by
%\begin{equation}
%  \displaystyle \prod_{i=1}^{N}[p_{mem_i}p(a_i)]^{P_{dsph_i}}[p_{non_i}(1-p(a_i))]^{1-P_{dsph_i}}.
%  \label{eq:emlikelihood}
%\end{equation}
We use an iterative, expectation-maximization (EM) algorithm \citep{sen07} to recover maximum-likelihood estimates of the Gaussian means and variances as well as the individual membership probabilities (see Walker et al.\ in prep. for full details of the algorithm).  Black markers in Figure \ref{fig:members1} identify the most likely interlopers ($\hat{P}_{dsph_i} < 0.01$), while colored markers signify larger $\hat{P}_{dsph_i}$ (see caption to Figure \ref{fig:members1}).  In subsequent calculations we weigh each data point by $\hat{P}_{dsph_i}$.

Middle sub-panels in Figure \ref{fig:members1} plot velocity against angular distance from the dSph center.  Bottom sub-panels indicate the overall velocity distribution; red lines give the expected distributions of interloper velocities, per the Besan{\c c}on model.  For dSphs observed with a high degree of contamination (Carina, Sextans), we find excellent agreement between the predicted and observed velocity distributions of foreground stars.  Table \ref{tab:table} lists for each dSph the numbers of newly observed stars, total stars after combining with published data, and member stars $N_{dsph}=\sum_{i=1}^{N_{tot}} \hat{P}_{dsph_i}$.

\section{Velocity Dispersion and NFW Profiles}

We estimate line-of-sight velocity dispersion profiles after dividing each sample into $\sim \sqrt{N_{dsph}}$ bins according to projected distance from the dSph center.  For a given sample we define bins (circular annuli) such that all contain approximately equal numbers of dSph members.  Thus the number of stars, including interlopers, in each bin may vary but for all bins, $\Sigma_{i=1}^{N_{bin}}\hat{P}_{dsph_i} \sim \sqrt{N_{dsph}}$.  We use a Gaussian maximum-likelihood method (see \citealt{walker06a}) to estimate the velocity dispersion within each bin.  
\begin{figure*}
  \epsscale{0.97}
  \plotone{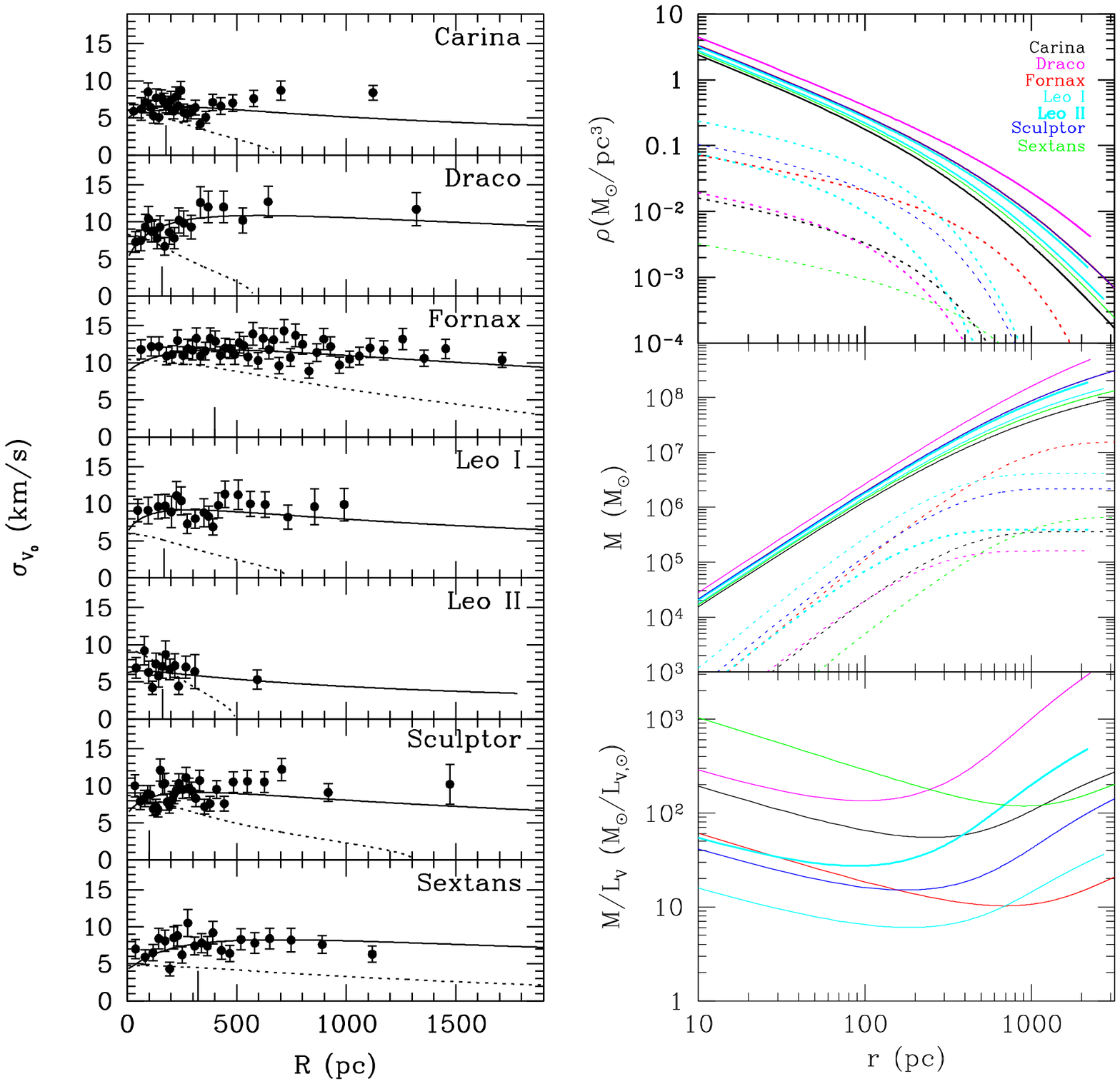}
  \caption{\textit{left:} Projected velocity dispersion profiles for seven Milky Way dSph satellites.  Overplotted are profiles corresponding to mass-follows light \citep{king62} models (dashed lines; these fall to zero at the nominal ``edge'' of stellar distribution), and best-fitting NFW profiles that assume constant velocity anisotropy.  Short, vertical lines indicate luminous core radii (IH95).  Distance moduli are adopted from \citet{mateo98}.  \textit{right:} Solid lines represent density, mass and $M/L$ profiles corresponding to best-fitting NFW profiles.  Dotted lines in the top and middle panels are baryonic density and mass profiles, respectively, following from the assumption that the stellar component (assumed to have M/L=1) has exponentially falling density with scale length given by IH95.}
  \label{fig:profiles}
\end{figure*}

Left-hand panels Figure \ref{fig:profiles} display the resulting velocity dispersion profiles, which generally are flat.  The outer profile of Draco shows no evidence for a rapidly falling dispersion, contrary to evidence presented by \citet{wilkinson04}, but consistent with the result of \citet{munoz05short}\footnote{We have not included the unpublished data of \citet{wilkinson04} or \citet{munoz05short} in our calculations of the velocity dispersion profiles of Draco.}.  In fact the outer profiles of Draco, Carina and perhaps Sculptor show gently rising dispersions.  While it is likely that at least in Carina this behavior is associated with the onset of tidal effects \citep{munoz06short}, \citet{mcconnachie06} point out that the tendency of some dSphs to have systematically smaller velocity dispersions near their centers is perhaps the result of distinct and poorly mixed stellar populations \citep{tolstoy04short,battaglia06short,ibata06}.  Either explanation complicates a thorough kinematic analysis; in the present, simplified analysis we assume all stars belong to a single population in virial equilibrium.

Dashed lines in Figure \ref{fig:profiles} are velocity dispersion profiles calculated for single-component King models \citep{king62} conventionally used to characterize dSph surface brightness profiles.  The adopted King models are those fit by \citet[hereafter IH95]{ih95} and normalized to match the central velocity dispersion.  Their systematic failure to predict flat profiles removes any doubt that (Newtonian) mass-follows-light, equilibrium models provide a poor description of dSph kinematics.  

In forthcoming papers we examine evidence for tidal effects and explore the range of dark matter density profiles that are consistent with the dSph velocity data.  Here we simply show that dark matter halos having the NFW density profile can fit the data reasonably well.  Under the assumptions of 1) spherical symmetry, 2) dynamic equilibrium, 3) radially constant velocity anisotropy $\beta\equiv 1-\sigma_{\theta}^2/\sigma_r^2$, and that 4) dSph surface brightness declines exponentially with radius, with central value and scale length adopted from IH95, we use the Jeans equation (e.g., Eq. A15 of \citealt{mamon05}) to calculate projected velocity dispersion profiles expected for NFW halos over a range of plausible virial masses.  Assuming the relationship between $M_{vir}$ and halo concentration found by \citet[see also \citealt{mamon05,koch07a}]{jing00}, we fit to the empirical velocity dispersion profiles using just two free parameters, $M_{vir}$ and $\beta$.  Solid lines in the left-hand panels of Figure \ref{fig:profiles} indicate the best-fitting NFW models, and Table \ref{tab:table} lists the associated parameters, including $M_{r_{max}}$, the enclosed (dark plus luminous) mass at the outermost profile point.  The right-hand panels of Figure \ref{fig:profiles} display mass-density, mass and $M/L$ profiles corresponding to the best-fitting NFW models.

\begin{deluxetable}{lrrrrrrrrrr}
\tabletypesize{\scriptsize}
\tablewidth{0pc}
\tablecaption{ Summary of Velocity Samples and NFW Parameters}
\tablehead{\colhead{Galaxy}&\colhead{$N_{new}$}&\colhead{$N_{tot}$}&\colhead{$N_{dsph}$}&\colhead{$\beta$}&\colhead{$M_{vir}; M_{r_{max}}; M_{600}$}\\
\colhead{}&\colhead{}&\colhead{}&\colhead{}&\colhead{}&\colhead{($10^7M_{\sun}$)}
  }
  \startdata
  Carina &1833& 2567& 899&-0.5&20; 3.5; 2.0\\
  Draco &512& 738& 413&-1&400; 9.0; 6.9\\
  Fornax &1924& 2085& 2008&-0.5&100; 18; 4.6\\
  Leo I &371& 483& 416&-0.5&100; 7.3; 4.5\\
  Leo II &128& 264& 213&0&40; 4.3; 2.8\\
  Sculptor &1089& 1214&1091&-0.5&100; 8.2; 4.3\\
  Sextans &947& 1032& 504&-2&30; 5.4; 2.5
  \enddata
  \label{tab:table}
%  \tablenotetext{a}{reference: \citet{mateo98}}
%  \tablenotetext{N04}{from \citet{navarro04} density profile}
%  \tablenotetext{W05}{from non-parametric mass estimation technique of \citet{wang05}}
\end{deluxetable}

\citet{strigari07} show that dSph masses within $r \leq 600$ pc ($M_{600}$) are constrained robustly by kinematic data, irrespective of the assumed form of the density profile.  This radius is convenient for practical reasons in our analysis as it covers the largest region common to all the data sets and is sufficiently small that one expects tidal effects inside this radius to be negligible \citep{read06}.  Integrating the density profiles for the stellar (for which we assume $M/L=1$) and dark components, we find that $M_{600}$ ranges from $(2-7) \times 10^7 M_{\odot}$ among the seven dSphs (Table \ref{tab:table}).  This narrow range is consistent with the notion that these dSphs, while exhibiting order-of-magnitude variability in luminosity, occupy dark halos of similar mass \citep{mateo93}.  

%\section{Discussion}
%
%We have assumed for simplicity that tides are dynamically insignificant.  A growing body of evidence suggests that some of these systems are tidally perturbed in their outer regions \citep{munoz06short,sohn06short,mateo07}, which can inflate stellar velocity dispersions and resulting mass estimates above equilibrium values.  A proper consideration of tides may reduce the derived masses and might mitigate the outlier status of Draco, which, of the sampled dSphs is nearest to the MW.  However, even in the few well-studied cases, tides do not eliminate the requirement of a significant DM component \citep{klessen03,mateo07,munoz06short,sohn06short}.  Furthermore, in light of the flat empirical profiles, the recent N-body simulations by \citet{read06} indicate it is unlikely that tides significantly affect the dynamics of dSphs at radii $r \leq 1$ kpc.  In a forthcoming paper we test for kinematic evidence of tidal effects in each of our samples.  
  
We thank the staff at the Las Campanas and MMT Observatories.  This work is supported by NSF Grants AST 05-07453, AST 02-06081, and AST 94-13847.

\bibliography{ref}

\end{document}